# Ariadne: PyTorch Library for Particle Track Reconstruction Using Deep Learning

Pavel Goncharov[1,a)], Egor Schavelev[2], Anastasia Nikolskaya[2], Gennady Ososkov[1]

[1]*Joint Institute for Nuclear Research, 6 Joliot-Curie street, 141980, Dubna, Moscow region, Russia*
[2]*Saint Petersburg State University, 7/9 Universitetskaya Emb., 199034, Saint Petersburg, Russia*

[a)]Corresponding author: kaliostrogoblin3@gmail.com

**Abstract.** Particle tracking is a fundamental part of the event analysis in high energy and nuclear physics (HENP). Events multiplicity increases each year along with the drastic growth of the experimental data which modern HENP detectors produce, so the classical tracking algorithms such as the well-known Kalman filter cannot satisfy speed and scaling requirements. At the same time, breakthroughs in the study of deep learning open an opportunity for the application of high-performance deep neural networks for solving tracking problems in a dense environment of experiments with heavy ions. However, there are no well-documented software libraries for deep learning track reconstruction yet. We introduce Ariadne, the first open-source library for particle tracking based on the PyTorch deep learning framework. The goal of our library is to provide a simple interface that allows one to prepare train and test datasets and to train and evaluate one of the deep tracking models implemented in the library on the data from your specific experiment. The user experience is greatly facilitated because of the system of gin-configurations. The modular structure of the library and abstract classes let the user develop his data processing pipeline and deep tracking model easily. The proposed library is open-source to facilitate academic research in the field of particle tracking based on deep learning.

## INTRODUCTION

Track building process or tracking is a key part of event reconstruction in High Energy and Nuclear Physics (HENP) analysis. It consists of grouping a set of signals registered with the help of a tracking detector by their belonging to some particle track. These signals are named hits and the hits referring to one of the tracks must obey a condition of lying on the smooth curve that has a form close to the spatial helix.

There are a lot of difficulties that interfere with tracking. The initial number of tracks is unknown. The primary vertex – the point of interaction of ion beams (in collider) and ion beam and fixed target (fixed target experiment) is unknown. Different tracks may have different lengths, e.g., some tracks had a small impulse and left the sensitive detector area before flying through all detector's coordinate planes or so-called stations. Each type of tracking system has its drawbacks. For example, if we are considering microstrip GEM detectors a lot of fake hits appear along with the real ones because of the spurious strip crossings and the amount of these fakes is quadratically greater than the number of true hits [1]. More and more experiments with high luminosity and events multiplicity are developed, the most famous are Large Hadron Collider (LHC) [2] and NICA project [3]. By 2024, the fourth launch is planned at the LHC, physicists expect about 10 thousand tracks in the event. All the above-mentioned obstacles make the tracking procedure extremely difficult.

On the other hand, a set of classical tracking algorithms already exist, most of them are based on Kalman Filter (KF) [4] and achieve a very good efficiency but at the cost of processing speed. All these methods have several disadvantages associated with the KF computational complexity and significant difficulties in the implementation of parallel computational schemes for building tracks. The disadvantages of KF are especially pronounced in experiments with a huge multiplicity of tracks. The problem is that although KF has an acceptable efficiency but it is very slow and together with the speed of data production in modern experiments it becomes impossible to process incoming data online or store all this information on hard drives. So, a new faster approach to particle tracking is now a real challenge.

In the same time, deep learning methods made a revolution in many fields due to their ability to model complex nonlinear dependencies of data. Besides, deep learning networks operate with linear algebra as the forward and backward pass of any artificial neural network is nothing less than a sequence of matrix multiplications alternating applying of nonlinearities. Linear algebra operations can be easily parallelized on modern high-performance processors. These properties of deep learning models make them very attractive to be used in the tracking, and the confirmation of this statement is that there are already several works on the application of deep learning to solving the tracking problem [5, 1, 6, 7].

Despite all progress in implementing deep learning to the event reconstruction, there are no well-documented software libraries for deep learning track reconstruction yet. Existing toolkits and repositories don't meet many conditions. There is a Toolkit for Multivariate Data Analysis with ROOT (TMVA) [8] which includes many machine learning methods and is compatible with ROOT framework for HENP data analysis, but it has several important drawbacks. The functionality of TMVA is very limited, it supports only base methods of artificial neural networks and can't be compared with any modern deep learning frameworks such as PyTorch [9]. Also opened manuals for TMVA is rather outdated, the published user guide dates back to 2007. And the main difficulty is that the user must know C++ syntax which is subjectively complex.

Another contribution to the deep learning tracking is a repository of Exa.TrkX project [5]. This repository supports the graph neural network (GNN) model was specially designed to solve the Kaggle TrackML challenge [10]. The authors report a great relative efficiency of over 95% for all particles. The problem is that this repository is very specialized as it applies only to TrackML data and can't be simply applied to any other experiment. Also, our experience has shown that direct application of GNNs to the GEM detectors leads to a very bad efficiency of tracking [6].

To address these gaps, we introduce Ariadne [11], the first open-source library for particle tracking based on the PyTorch deep learning framework. We named it Ariadne in the fame of Ariadne's thread which means the path leading to the goal in difficult conditions as in the tracking process. Our main contributions are as following. We propose an open-source library for particle tracking using deep learning methods. The library has a modular structure and abstract classes simplifying the process of developing a custom model and data processing pipeline for your data. We implement all our best models in Ariadne with the ability to train them on an arbitrary tracking task. The library includes a system of configuration files for fast reproducing of experiments with 100% deterministic behavior. Ariadne has a tool for events visualization with the ability to highlight separate tracks and apply filtration transforms.

Our library is still in development, but already can be used to facilitate academic research in the field of particle tracking based on deep learning.

## ENVIRONMENT OVERVIEW

We have chosen PyTorch [9] as a base deep learning framework for our library because it is open source, flexible, has a great variety of neural network layers, activation functions, and optimizers. PyTorch allows users to train deep learning models in a GPU or TPU environment, supports automatic half-precision. The computation graphs build in PyTorch dynamically that simplifies the process of research and development and makes the process of debugging clearer.

Nevertheless, PyTorch can be named as a low-level framework, because the user has to manually program the overall training process and it becomes harder when some options such as logging metrics, distributed training, etc. are required. Fortunately, the ecosystem of PyTorch includes a great number of libraries were written on top of PyTorch and we have chosen PyTorch Lightning [13] as a fundament of organizing the training process in Ariadne. We created a special class *TrainModel* which is a Lightning module to extend the functionality of simple PyTorch code with Lightning features: full reproducibility, checkpointing, callbacks, metrics logging, multi-GPU training, TPU training, learning rate schedulers, batch size optimization, etc.

To allow the users to reproduce our experiments and set up their own easily we introduce a structure of configuration files, consisting of train configurations, data preparation configurations, and inference/evaluation configurations. Standard Python *ArgumentParser* is not very flexible so we address Gin Config [14] to this task. Gin Config opens an opportunity to specify the whole classes with their arguments inside a configuration file, so in the future, all the users of our library will be able to "program" any training from scratch using Ariadne modules and such an approach requires no knowledge of Python.

Almost each program library has some hardware and software requirements. When the user wants to make use of any library or framework, he should follow the installation guide and set up all required dependencies. In order to simplify the process of setting up the working environment Python has the mechanism of virtual environments and its own package manager named pip. The problem with pip is that it does not control the conflicts between different packages and versions and not build the binaries for a particular system, e.g., Windows. It is a well-known problem and there are even several repositories of precompiled binaries of many Python packages. In order to solve the problem of setting the environment to work with Ariadne, we prefer to use conda package management system and environment management system [15]. We created two configurations of conda environments – for GPU and CPU usage and place them in the root of Ariadne repository. So now the user doesn't need to build the environment, download dependencies, and manage conflicts between libraries versions by himself – conda will manage all for him.

As we are working on the base of Laboratory of Information Technologies, we want to be able to run all our scripts in the HybriLIT's environment and especially on the GOVORUN supercomputer [16]. We added several scripts in the 'scripts/hydra' path for that purpose [11].

The full process of setting up the environment and running scripts on HybriLIT platform is described in the README file of Ariadne repository [11].

## COMPONENTS

Following the scheme of almost any deep learning pipeline we split up all functionality into three main concept blocks: data preparation step, training step, and inference, or the part that is responsible for using the already trained model in test time or real application. Each stage has its semantic components. We named them as Processor, Parser, TrainModel, DataLoader, etc. Most of them are illustrated in Fig. 1.

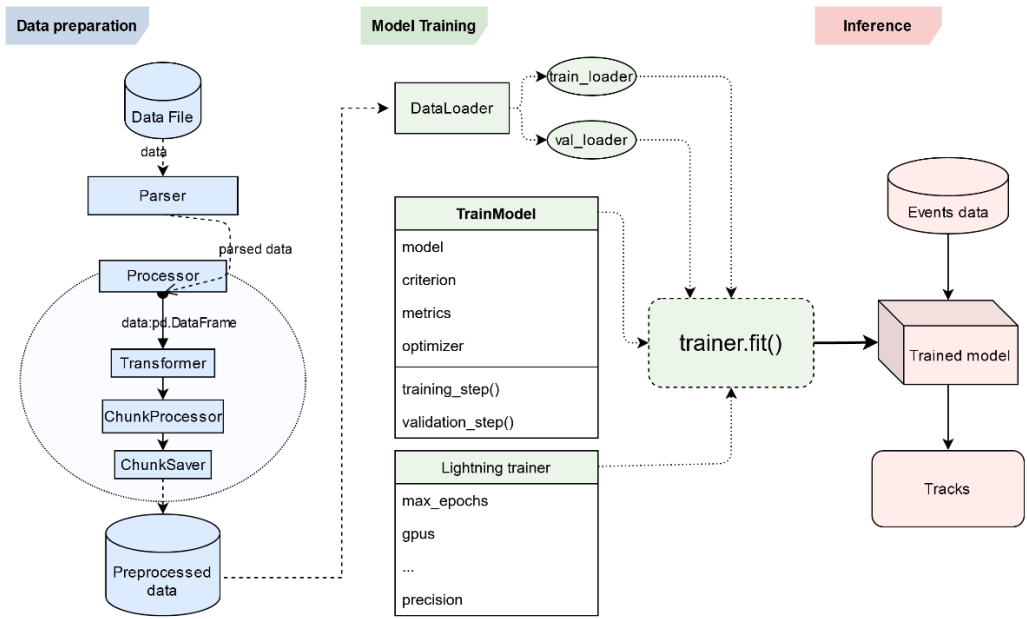

**FIGURE 1.** The scheme of components of Ariadne

For the data preparation step, we introduce transformations that allow us to translate coordinates into another system, e.g., cartesian to cylindrical, filter out inapplicable tracks or events, normalize and rescale features to be in a unit variance. At the moment, we haves plenty of transformations: StandardScale, MinMaxScale, Normalize, ConstraintsNormalize, DropShort, DropSpinningTracks, DropFakes, ToCylindrical, ToCartesian, etc. Inspired by the PyTorch torchvision transforms we created a special Compose class that can combine multiple transformations in one block for convenience.

Every model should implement its Processor by realizing the class inherited from DataProcessor. Processor is responsible for data preparation for training and in the future will be a part of inference pipeline to reduce the amount of code. Processor operates with data divided into chunks. Each chunk may be either an event or a track-candidate. The logic of splitting data into chunks contained in the function "generate_chunks_iterable" should be implemented for each class inherited from DataProcessor. After the chunks being generated, they are processed using transforms and then can be combined and saved on disk as a complete dataset with the help of functions "postprocess_chunks" and "save_on_disk" respectively. Processor is used before the training to create the train and test dataset.

As for the training part, we highlighted several classes. The first and almost the main block is the TrainModel. TrainModel is a Lightning module to extend the functionality of simple PyTorch code with Lightning features. It has an attribute "model" – PyTorch "nn.Module" class which should be used in both training and inference. Every custom Ariadne model should be implemented as a class inherited from "nn.Module", but the TrainModel class of our Ariadne library is shared across all existing and future models. This was made for convenience. The TrainModel class encapsulates all the necessary training routines such as loss and metrics calculation, statistics logging, etc.

Except for the model class, each algorithm should provide its metrics, criterion, and dataset with DataLoader. At this moment, metrics are simple Python functions that take as input model's predictions and the target for optimization. Criterion is a class that implements loss function which will be used for the model's training. Dataset is a PyTorch dataset class inherited from "torch.utils.data.Dataset". Each model implements its dataset, metrics, and criterion.

At the moment of writing this article, the inference part of the library is still in progress, but it is already possible to foresee that it will have a component for combining the data preprocessing part and model application to the events data. Each model should take the events as input in the same format as for other models and output a list of tracks represented by a sequence of IDs of hits. After that several metrics can be applied to analyze the efficiency of the tracking algorithm such as dependence of efficiency on the transverse momentum of particles, so it would be possible to compare with other tracking methods.

The library has a simple structure. Each tracking method can be added as a separate module to the Ariadne library. The files of the module should be placed in the source folder of the library and the directory with the module should contain at least 6 files with a model, processor, metrics, loss, dataset, and the file for module initialization in Python, so-called "__init__" file. The main scripts for the data preparation step and training are located in the root directory of the repository. The directory named "resources" includes required resources for training such as training configuration files and source examples for unit testing. "test" folder is a place for unit tests, "notebooks" includes some exploratory data analysis, the directory named "visual" contains the visualization tool (see the Section "Visualization tool") for the events data and there are some other folders that are not interesting in terms of this paper.

Ariadne currently, although is in the active use, but at the same time, it is under the stage of continued development and some blocks may be replaced or renamed, but the main conception will remain the same.

## VISUALIZATION TOOL

Successful solving modern Machine Learning (ML) tasks demands a careful examination of the data. Such would not have been possible without a proper visualization tool. This tool should be fast in terms of performance, run on almost any machine without any complicated installation steps or downloads. That is why we decided to develop our visualization framework on top of PyQt [17] and Vispy [18] packages. We chose the PyQt library because it can run almost on any operating system. But one of the main goals of such an instrument is to draw data rapidly. In oppose to Matplotlib [19], which is reasonably slow when working with a lot of data but can produce beautiful pictures, Vispy library allows visualizing hundreds of thousands of points in 30 frames per second.

The idea of our visualization toolkit is to be able to display events from modern detectors, which have more than 1000 tracks in a single event almost immediately. On top of that, we propose a novel mechanism for researching such complex data: just-in-time (JIT) data transforms. Users can apply any data transformation in an interactive console and see the data without restarting the app. JIT transforms are saving a lot of researcher's time and, as a result, speeds up greatly the data investigation and research phase of any ML experiment. The tool is not feature-complete yet, but it is already can display the 3D picture of a very dense event. In an interactive mode, the user can select the camera, rotate and zoom the 3D scene in real-time 30 FPS on any modern laptop or PC.

## CONCLUSION

In this paper we introduced Ariadne – the first library for deep learning tracking on Python. This library is highly configurable, it has a simple interface to implement new deep learning models for almost any type of event data including collider and fixed-target experiments. Ariadne has several scripts that will help the user to make data preparation and run training. As the library uses PyTorch Lightning interfaces as the main skeleton for training it provides metrics logging, multiprocessing for data preparation, multi-GPU training, etc. Ariadne is open source and fully deterministic so it is very easy to reproduce some experiment and reproduce the results of any scientific paper which will use Ariadne. With the help of Ariadne, we have already trained two models – TrackNETv2 [1] and GraphNet [6] on the BES-III data [12] and obtained the mean efficiency of over 95% for both of the models.

At the moment we are working on code refactoring and development of the interfaces for inference and testing, also the extension of documentation is required.

## ACKNOWLEDGMENTS

The calculations were carried out on the basis of the HybriLIT heterogeneous computing platform (LIT, JINR) [16].

The reported study was funded by RFBR according to the research project № 18-02-40101.